\documentclass[aps,prl,twocolumn,showpacs,superscriptaddress,10pt]{revtex4}

\usepackage{color}
\usepackage{graphicx}
\usepackage{dcolumn}
\usepackage{bm}
\usepackage{amsbsy}
\usepackage{amsmath,amssymb}
\usepackage[hypertex]{hyperref}

\newcommand{\nc}{\newcommand}
\nc{\rnc}{\renewcommand}
\nc{\A}{\mb{\boldmath$A$}}
\nc{\be}{\begin{eqnarray}}
\nc{\bea}{\begin{eqnarray}}
\nc{\bean}{\begin{eqnarray*}}
\nc{\bra}[1]{\langle #1 |}
\nc{\ee}{\end{eqnarray}}
\nc{\eea}{\end{eqnarray}}
\nc{\eean}{\end{eqnarray*}}
\nc{\h}{\hspace{1pt}}
\nc{\ket}[1]{| #1 \rangle}
\nc{\mb}{\mbox}
\nc{\mol}{_\mathrm{mol}}
\nc{\nab}{\bm{\nabla}}
\nc{\rr}{\mb{\boldmath$r$}}
\nc{\Tr}{\mathrm{Tr}}
\nc{\va}{\mb{\boldmath$a$}}
\nc{\vB}{\mb{\boldmath$B$}}
\nc{\vE}{\mb{\boldmath$E$}}
\nc{\vg}{\mb{\boldmath$g$}}
\nc{\vH}{\mb{\boldmath$H$}}
\nc{\vj}{\mb{\boldmath$j$}}
\nc{\vk}{\mb{\textbf k}}
\nc{\vM}{\mb{\boldmath$M$}}
\nc{\vn}{\mb{\boldmath$n$}}
\nc{\vp}{\mb{\boldmath$p$}}
\nc{\vP}{\mb{\boldmath$P$}}
\nc{\vpi}{\mb{\boldmath$\pi$}}
\nc{\vq}{\mb{\boldmath$q$}}
\nc{\vR}{\mb{\boldmath$R$}}
\nc{\vs}{\mb{\boldmath$\sigma$}}
\nc{\vS}{\msb{\boldmath$S$}}
\nc{\vt}{\mb{\boldmath$\tau$}}
\nc{\vz}{\hat {\mb{\textbf z}}}
\nc{\x}{\mb{\boldmath$x$}}
\nc{\X}{\sf x}
\rnc{\vec}[1]{\mathbf{#1}}

\begin{document}

\title{Mechanisms of enhanced orbital dia- and paramagnetism: \\ Application to
the Rashba semiconductor BiTeI}

\author{G.~A.~H.~Schober}
\email{g.schober@thphys.uni-heidelberg.de}
\affiliation{Department of Applied Physics, University of Tokyo, Tokyo 113-8656,
Japan}
\affiliation{Institute for Theoretical Physics, University of Heidelberg,
D-69120 Heidelberg, Germany}
\affiliation{Cross-Correlated Materials Research Group (CMRG), ASI, RIKEN, Wako
351-0198, Japan}

\author{H.~Murakawa}
\affiliation{Correlated Electron Research Group (CERG), ASI, RIKEN, Wako
351-0198, Japan}

\author{M.~S.~Bahramy}
\affiliation{Correlated Electron Research Group (CERG), ASI, RIKEN, Wako
351-0198, Japan}

\author{R.~Arita}
\affiliation{Department of Applied Physics, University of Tokyo, Tokyo 113-8656,
Japan}
\affiliation{Correlated Electron Research Group (CERG), ASI, RIKEN, Wako
351-0198, Japan}

\author{Y.~Kaneko}
\affiliation{Multiferroics Project, Exploratory Research for Advanced Technology
(ERATO), Japan Science and Technology Agency (JST), c/o Department of Applied
Physics, University of Tokyo, Tokyo 113-8656, Japan}

\author{Y.~Tokura}
\affiliation{Department of Applied Physics, University of Tokyo, Tokyo 113-8656,
Japan}
\affiliation{Cross-Correlated Materials Research Group (CMRG), ASI, RIKEN, Wako
351-0198, Japan}
\affiliation{Correlated Electron Research Group (CERG), ASI, RIKEN, Wako
351-0198, Japan}
\affiliation{Multiferroics Project, Exploratory Research for Advanced Technology
(ERATO), Japan Science and Technology Agency (JST), c/o Department of Applied
Physics, University of Tokyo, Tokyo 113-8656, Japan}

\author{N.~Nagaosa}
\email{nagaosa@ap.t.u-tokyo.ac.jp}
\affiliation{Department of Applied Physics, University of Tokyo, Tokyo 113-8656,
Japan}
\affiliation{Cross-Correlated Materials Research Group (CMRG), ASI, RIKEN, Wako
351-0198, Japan}
\affiliation{Correlated Electron Research Group (CERG), ASI, RIKEN, Wako
351-0198, Japan}

\date{\today}

\begin{abstract}
We study the magnetic susceptibility of a layered semiconductor BiTeI with giant
Rashba spin splitting both theoretically and experimentally to explore its
orbital magnetism. Apart from the core contributions, a large
temperature-dependent diamagnetic susceptibility is observed when the Fermi
energy $E_F$ is near the crossing point
of the Rashba spin-split conduction bands at the time-reversal symmetry point A.
On the other hand, when $E_F$ is below this band crossing the susceptibility
turns to be paramagnetic. These
features are consistent with first-principles calculations, which also predict
an enhanced orbital magnetic susceptibility with both positive and negative
signs as a function of $E_F$ due to band (anti)crossings. Based on these
observations, we propose two mechanisms for the enhanced paramagnetic orbital
susceptibility.
\end{abstract}


\maketitle

The magnetic susceptibility $\chi$ is one of the most fundamental physical
quantities, revealing invaluable information about the spin and orbital states
of solid materials~\cite{Mermin}. The sign of $\chi$, depending sensitively on
the mechanism of the magnetization, plays a crucial role in identifying the
magnetic properties of bulk systems. From this viewpoint, solid materials are
roughly classified into the following three categories~\cite{Mermin}: (i) band
insulators with weak temperature-independent diamagnetism $\chi<0$, (ii) metals
with temperature-independent Pauli paramagnetic susceptibility $\chi_P>0$ and
temperature-independent Landau diamagnetic susceptibility $\chi_L<0$ well below
the Fermi temperature $T \ll T_F= E_F/k_B$, and (iii) magnetic materials with
local moments with strongly temperature-dependent Curie paramagnetic
susceptibility $\chi(T) \propto 1/T$ or Curie-Weiss law $\chi(T) \propto
1/(T-T_0)$ corresponding to ferromagnetic ($T_0>0$) or antiferromagnetic
($T_0<0$) interactions between the magnetic moments. When an electron
configuration with $J=0$ ($J$: total angular momentum) is realized for a
magnetic ion, e.g., in the case of four $d$-electrons, Larmor diamagnetism and
van-Vleck paramagnetism compete to determine the sign of $\chi$.

Usually, the orbital motion of electrons leads to a diamagnetic $\chi$, as in
the case of Larmor and Landau diamagnetism. This is due to the cyclotron motion
induced by the Lorentz force, which tends to reduce the action of the external
magnetic field.
When the band gap is reduced or fully closed, this diamagnetic susceptibility is
considerably enhanced as in the cases of bismuth~\cite{nolting} and
graphene~\cite{Neto}. Especially in graphene, due to the presence of gapless
two-dimensional Dirac fermions, $\chi$ diverges as {$\propto -1/T$} as $T \to
0$~\cite{Neto}. Despite this general tendency, there is no proof that the
orbital motion always favors diamagnetism, in particular in the presence of
strong spin-orbit interaction (SOI). Hence one cannot exclude the possibility
that orbital paramagnetism may be realized in band insulators
and metals. So far, however, this possibility has been considered
by only a few theoretical studies~\cite{Kubo,Rashba,Vignale,Bruder,Principi}.
For example, Vignale~\cite{Vignale} showed that a two-dimensional (2D) electron gas
in a periodic potential exhibits orbital paramagnetism when
the Fermi energy is close to a saddle point of the band structure.
Boiko and Rashba~\cite{Rashba} predicted orbital magnetism in the Rashba model,
for which we will provide the first material realization and its further theoretical
analysis below. For band insulators, it has so far been an open question whether orbital
paramagnetism can be realized or not.

In this paper, we address the mechanisms
of orbital dia- and paramagnetism in semiconductors by studying BiTeI, a
narrow-gap polar semiconductor with a noncentrosymmetic layered
structure. Due to the polarity of the bulk material and the presence of Bi atoms
with large atomic SOI, BiTeI exhibits an extraordinarily large  bulk Rashba spin
splitting (RSS), astonishingly reaching {$\simeq 400$ meV} at the bottom of the
conduction band as observed in the  recent angle-resolved photoemission
spectroscopy experiment by Ishizaka~\textit{et~al.}~\cite{Ishizaka}. Further
studies based on optical spectroscopy~\cite{lee2011}, first-principles
calculations and group theoretical analysis~\cite{Bahramy} have confirmed that
this giant RSS is indeed a bulk property of BiTeI, and revealed that in addition
to the bottom conduction bands (BCB's), the top valence bands (TVB's) of BiTeI
are subject to a comparable RSS. These unique features  make BiTeI an ideal
medium to explore different aspects of orbital magnetism in a band insulator
system. Below we theoretically predict and experientally confirm that in BiTeI
both an enhanced orbital diamagnetism and an unconventional orbital
paramagnetism can be realized by tuning the Fermi energy $E_F$. This intriguing
behavior is shown to be a direct consequence of the RSS.

\begin{figure}[t]
        \begin{center}
                \includegraphics[width=1.0 \linewidth]{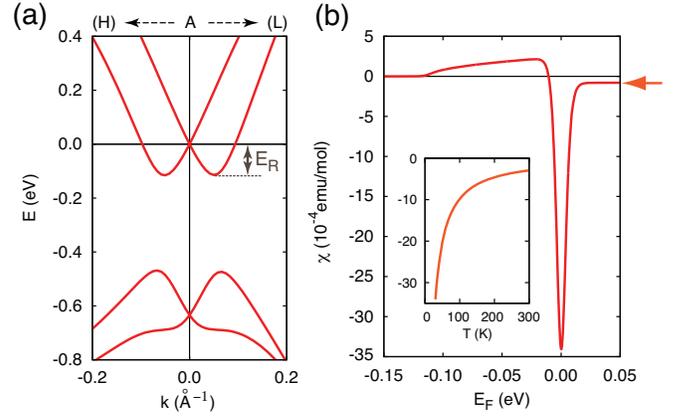}
                \caption{(Color online) (a) Band dispersion of BiTeI near
point~$A$ towards the $H$ and $L$ directions. The crossing point of the
conduction bands is set at zero energy. (b) Orbital magnetic susceptibility
obtained from a 2D Rashba model, which reproduces the dispersion of the BCB's in
BiTeI around point~$A$. The main panel shows the Fermi energy dependence
of~$\chi$ at $T = 30$~K. The arrow marks the value of the Landau diamagnetic
susceptibility for free electrons with mass $m^*$ in 2D (see the related
discussion). The inset shows the temperature dependence of the diamagnetic peak
value at $E_F = 0$, diverging as $-1/T$.\label{1}}
        \end{center}
\end{figure}

As described in Refs.~\onlinecite{Ishizaka,Bahramy}, BiTeI has a minimum band
gap not at the Brillouin zone (BZ) center, but around the hexagonal face center
of the BZ, referred to as point~$A$, as shown in Figure~\ref{1}(a). Due to the
strong covalency and ionicity of Bi-Te and Bi-I bonds, respectively, the BCB's
are dominated by Bi-$6p$ states, whereas the TVB's are predominantly of Te-$5p$
character with a partial contribution from I-$5p$ states. The huge SOI of Bi
accompanied by a strong negative crystal field splitting of TVB's leads BCB's
and TVB's to be symmetrically of the same character, and hence to be strongly
coupled with each other via a quasi-two-dimensional RSS Hamiltonian. Near
point~$A$, this can be described as~\cite{Bahramy}
\begin{equation}
	H_R = { \frac{{\mathbf p}^2}{2 m^*} } + \lambda {\mathbf e}_z \cdot
({\mathbf s} \times {\mathbf p}),
\end{equation}
where $m^*$ is the effective mass of the carriers, $\lambda$ is the Rashba
parameter, ${\mathbf e_z}$ is the unit vector in $z$ direction, which is the
direction of the potential gradient breaking the inversion symmetry, and
${\mathbf s}$ and ${\mathbf p}$ are the spin and momentum operators,
respectively. Note in particular the crossing of the BCB's
shown in Figure~\ref{1}(a), which is due to the Kramer's degeneracy at the
time-reversal symmetry point A.
As fitting parameters, $m^*$ and $\lambda$ can be
tuned~\cite{parameters} such that \,$H_R$ can properly reproduce the electronic
dispersion of BCB's around point~$A$, especially the momentum offset of the
conduction band minimum (CBM), $|\mathbf p| = \hbar k_0$ with $k_0 =
0.052$~\AA$^{-1}$, and the Rashba energy $E_R=113$~meV, where $E_R$ is the
energy of the conduction bands at their crossing point with respect to CBM.

To calculate the orbital magnetic susceptibility $\chi$ we employ the Fukuyama
formula~\cite{Fukuyama},
\begin{equation} \label{eqFukuyama}
        \chi(T) = \mu_0 \frac{N_A}{N} \frac{e^2 \hbar^2}{2} \h k_B T \sum_{\ell}
\sum_{\mathbf k} \Tr \h [ G_{\ell} v_x G_{\ell} v_y G_{\ell} v_x G_{\ell} v_y ],
\end{equation}
where $N_A$ is the Avogadro constant, $N$ the number of carriers, $G_{\ell} =
[i\omega_{\ell} + E_F - H]^{-1}$ the one-particle thermal Green's function with
Matsubara frequencies $\omega_{\ell} = (2\ell+1) \pi k_B T$,
$\ell\in \mathbb Z$, and $v_i = \partial H / \partial k_i$ $(i=x,y,z)$
the velocity operators. For the above 2D Rashba model,
Figure~\ref{1}(b) shows
the calculated dependence of $\chi$ on the Fermi energy $E_F$.
Near the band crossing point, where we take the origin of $E_F$, an enhanced
diamagnetic susceptibility is observed, which diverges as {$\chi \propto - 1/T$}
as in the case of graphene~\cite{Neto}. For $E_F\gg E_R$, $\chi$ approaches
$-A_{\mathrm{mol}}\h\mu_0 e^2/(12\pi m^*)$,
which is equivalent to the value of the Landau diamagnetic susceptibility for
free electrons with mass $m^*$ in 2D (here $A_{\mathrm{mol}}$ is the molar
surface area of one BiTeI layer). In addition, we find the paramagnetic orbital
susceptibility when $E_F$ is below the band crossing point, which is a unique
feature attributed to the Rashba spin splitting~\cite{Rashba}.

In the zero temperature limit, by evaluating the frequency sum in
the Fukuyama formula we have obtained an analytic formula for the orbital
susceptibility
of the 2D Rashba model,
\begin{equation} \label{eqFukuyama_II}
        \chi(T = 0) = \mu_0 A_{\mathrm{mol}} \frac{e^2}{m^*} \frac{1}{8\pi k_0}
        \int\limits_{E_-(k)\leq E_F \leq
E_+(k)}\hspace{-0.5cm}\frac{8k^2-3k_0^2}{8k^2} \, \mathrm d k.
\end{equation}
Here $k = |\vec k|$ denotes the radial momentum variable, while the angular
integration has already been performed;
$E_{\pm}(k) = \hbar^2 k^2/2m^* \pm \lambda k$ are the two branches of
the Rashba dispersion and $k_0 = m^* \lambda /\hbar^2$ the minimum of the lower
branch.
For small band fillings, i.\,e., $E_F \cong -E_R$, the momentum integral is
limited
to the region where $k \cong k_0$, hence $\chi$ is positive and approximately
given by
\begin{equation}
 \chi(T = 0) \cong \mu_0 A_{\mathrm{mol}} \frac{e^2}{m^*}
\frac{5}{32\pi}\sqrt{1+E_F/E_R}.
\end{equation}
On the other hand, if the Fermi energy is near the band crossing,
i.\,e., $E_F \cong 0$, the region of small $k$ gives the dominant contribution, which
gives rise to the negative divergence of the susceptibility as $\chi \propto -1/|E_F|$.~\cite{comment}

As a supporting argument for the orbital paramagnetism at
low band fillings, we have shown in the supplemental material that
the energy of the lowest Landau level of the 2D Rashba model
in a perpendicular magnetic field $\vec B$ is always lower than the minimum
energy~$-E_R$ in the absence of $\vec B$.
Accordingly, for sufficiently low filling of the bands, by applying the magnetic field
the total energy of the system is lowered, a sign for the
occurrence of paramagnetism.
Since the Landau-Peierls formula~\cite{LP}, which
takes into account only intraband contributions, gives always a negative value for the
orbital magnetic susceptibility, the positive $\chi$ can be attributed to the
interband contributions, which are allowed due to the RSS of BCB's. Also note
that this behavior is different from that in the 2D Dirac model describing
graphene, which shows only a diamagnetic orbital susceptibility.

To further corroborate these results as well as to explore the effects of all
energy
bands, we have performed  a detailed analysis based on first-principles density
functional theory (DFT). The orbital magnetic susceptibility of BiTeI
has been computed using the Fukuyama formula~\eqref{eqFukuyama}, with
the velocity operator $v_i$ ($i = x, y, z$) represented in  matrix form as
\begin{equation}
	v_{i,nm} = \bra n v_i({\mathbf k}) \ket m = \frac{1}{\hbar} \Big\langle
n \, \Big| \, \frac{\partial H({\mathbf k})}{\partial k_i} \, \Big| \, m
\Big\rangle,
\end{equation}
where $\ket n$ corresponds to the $n$th eigenstate of $ H(\mathbf k)$. The
numerical computation of $v_{i,nm}$ has been done using the approach of
Ref.~\onlinecite{wang}, whereby $H(\mathbf k)$ is described in the basis of
so-called maximally localized Wannier functions (MLWF's)~\cite{souza}. To fully
take into account the contributions of all possible energy states, a set of 18
MLWF's, spanning the 12 highest valence bands and 6~lowest conduction bands,
has been generated with the \textsc{wannier90} code~\cite{mostofi} by
postprocessing~\cite{kunes}
the relativistic DFT calculations,
initially done using Perdew-Burke-Ernzerhof exchange-correlation functional~\cite{pbe}
and the augmented plane wave plus local orbitals method as implemented in
the \textsc{wien2k} package~\cite{wien2k}. For these calculations, the muffin
tin radii are set to $R_{MT}=2.5$ Bohr for all the atoms and the maximum
modulus for the reciprocal vectors $K_{max}$ is chosen such that $R_{MT}K_{max}=7.0$.
The BZ is sampled  using a $20\times 20\times 20$  k-mesh and a full structural
optimization is performed until the magnitude of force on each ion is less than
0.1 mRy/Bohr (see Ref.~\onlinecite{Bahramy} for more details). For the evaluation
of the momentum integral in Eq.~\eqref{eqFukuyama}
we have used a $500 \times 500 \times 50$ k-mesh to sample BZ.
We have also calculated the Pauli
paramagnetic susceptibility $\chi_P$, leaving the orbital degree of freedom
unaffected. Namely we have added only the Zeeman coupling $-\mu_B
\boldsymbol{\sigma} \cdot \mathbf B$ to the Hamiltonian and calculated the spin
magnetization to obtain $\chi_P$. However, the contribution of $\chi_P$ is much
smaller than the orbital $\chi$ as described below.

\begin{figure}[ht]
	\begin{center}
		\includegraphics[width=0.9 \linewidth]{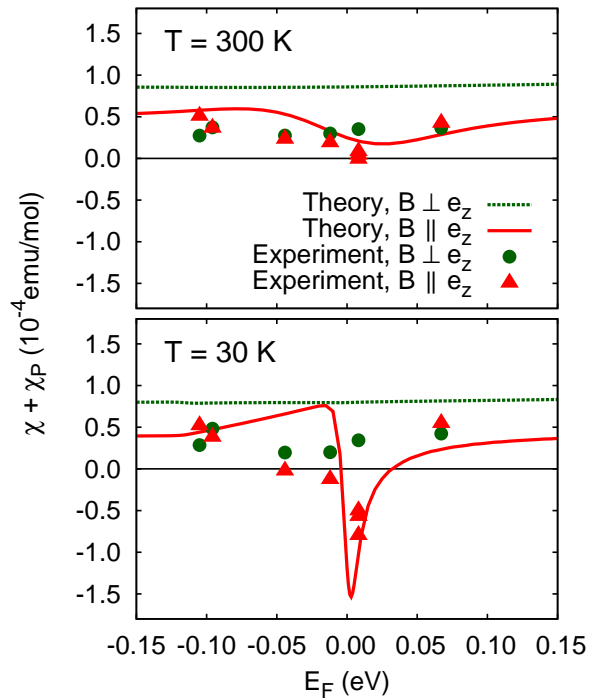}
		\caption{(Color online) Magnetic susceptibility of BiTeI as a
function of Fermi energy $E_F$, at two temperatures $T = 300$~K (upper panel)
and $T = 30$~K (lower panel). Solid lines represent the calculated
susceptibility including the Pauli contribution, and points show the
corresponding experimental data after subtracting the core contributions. $E_F =
0$ corresponds to the crossing point of the Rashba-split conduction
bands at point A.\label{2}}
	\end{center}
\end{figure}

Figure~\ref{2} shows the $E_F$ dependence of $\chi + \chi_P$ obtained by the
first-principles calculations for BiTeI taking into account the dispersion
along $k_z$ direction, where the origin of $E_F$ is taken at the crossing
point of BCB's at point A. The Pauli
contribution is an order of magnitude smaller than the orbital $\chi$, at least
in the considered range of $-0.15$~eV $\le E_F\le 0.15$~eV. This is due to the
small effective mass of BCB carriers within the hexagonal face of the BZ
(hereafter referred to as $ab$ plane), $m^* \simeq 0.2 m$~\cite{lee2011}. This
is analogous to the case of bismuth \cite{nolting} and can be understood by
considering a spin-degenerate two-dimensional parabolic band $E = \hbar^2 k^2 /
(2m^*)$. Then the orbital and Pauli susceptibility are given by $\chi = -(1/3)
\mu_0 N_A \h {\mu_B^*}^2 / E_F$ and $\chi_P = \mu_0 N_A \h {\mu_B}^2 / E_F$,
respectively, where $\mu_B^* = e\hbar / (2m^*)$, $\mu_B = e\hbar / (2m)$, and
$E_F = \hbar^2 k_F^2 / (2m^*)$ with the Fermi wave vector $k_F$. Since $m^*$
appears in $E_F$, but not in the spin-interaction term, i.e. in $\mu_B$, it
turns out that $\chi_P = -3 (m^*/m)^2 \chi \simeq -0.1 \chi$. Therefore we can
neglect $\chi_P$ and focus on $\chi$ below.

For $\mathbf{B}$ parallel to ${\mathbf e}_z$, near the band crossing point a
large temperature-dependent diamagnetic susceptibility is observed, while the
susceptibility turns to be paramagnetic and temperature-independent when $E_F$
is away from it. In contrast, if $\mathbf{B}$ is applied perpendicular to the
$z$ direction, $\chi$ is found to be always positive and almost insensitive to
$E_F$. The calculated results are in good agreement with the experiment, as
shown in Figure~\ref{2}.

For the experimental measurement of $\chi$, single crystals of BiTeI with
various charge carrier densities were grown by a Bridgman method or chemical
vapor transport. The obtained crystals have cleaved (001) planes. The charge
carrier density $n$ of each sample was determined by the Hall resistivity
measurement, and the corresponding Fermi energy $E_F$ was
deduced from the calculated relationship between $n$ and $E_F$
in the 18-band tight-binding model, as shown in Figure~3(a)
of Ref.~\onlinecite{lee2011}. The magnetization measurement was performed with use of the
vibrating-sample magnetometer. The sample holder for this was made of a plastic
straw tube, whose tiny diamagnetic signal was accurately subtracted.

To extract the orbital magnetic susceptibility from the experimental data, one
needs to subtract the Larmor susceptibly $\chi_L$ originating from the
individual ionic cores in BiTeI. Since in this material the ionic states of Bi,
Te and I ions are 3+, 2- and 1-, respectively, one can assume that they all have
closed shell ionic  configurations. Accordingly, for each ion $\chi_L$ can be
easily estimated as~\cite{Mermin}, $\chi_{L}=-\mu_0
N_A(e^2/6m)\sum_{i=n,l,j}Z_i\langle r^2 \rangle_{i}$, where $Z_i$ is the number
of electrons occupying state $i=\{n,l,j\}$. Performing a set of all-electron DFT
calculations using the \textsc{lda1.x} program~\cite{pwscf}, the spread
functions $\langle r^2 \rangle_{i}$ have been individually calculated for all
the occupied states of Bi$^{3+}$, Te$^{2-}$ and I$^{1-}$ (see Table 1 in the
{supplemental material}). Based on these calculations, the respective $\chi_L$
of Bi$^{3+}$, Te$^{2-}$ and I$^{1-}$ are found to be -32.5, -23.3 and -21.1 (all
in units of $10^{-6}$ emu/mol). We have accordingly subtracted these values from
our experimental data to deduce the orbital magnetic susceptibility.

\begin{figure}[h]
	\begin{center}
		\includegraphics[width=0.9 \linewidth]{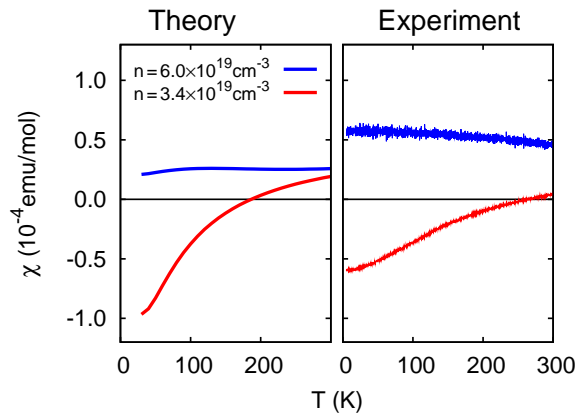}
		\caption{(Color online) Temperature dependence of the orbital
magnetic susceptibility of BiTeI for $\mathbf B \parallel \mathbf e_z$, obtained
theoretically (left panel) and experimentally (right panel) at two
characteristic carrier densities. For $n = 3.4\times10^{19}\mathrm{cm}^{-3}$ the
Fermi energy $E_F$ is close to the band crossing point, whereas for $n =
6.0\times10^{19}\mathrm{cm}^{-3}$ it is $\simeq 70$ meV above it.\label{3}}
	\end{center}
\end{figure}

In Figure~\ref{3}, we show the temperature dependence of the orbital magnetic
susceptibility for $\mathbf B \parallel \mathbf e_z$ at two characteristic Fermi
energies, one near the band crossing point ($E_F = 0$) and the other away from it ($E_F \simeq 70$ meV). In the
first case, a strong temperature dependence and sign change of $\chi$ from
paramagnetism to diamagnetism is observed as the temperature is lowered, while
in the latter case the susceptibility remains paramagnetic and almost
independent of temperature. This behavior is clearly understood in terms of the
temperature-dependent enhanced diamagnetic contribution near $E_F = 0$, which
adds to the almost temperature-independent paramagnetic interband contribution.
All these features are in excellent agreement with the experimental curves
measured at two different samples with corresponding carrier densities, as shown
in Fig~\ref{3}.

\begin{figure}[b]
        \begin{center}
                \includegraphics[width=0.9 \linewidth]{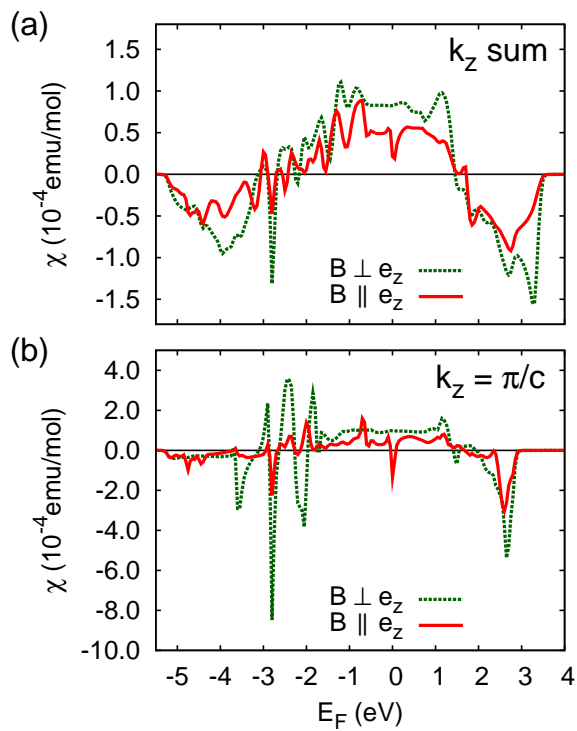}
                \caption{(Color online) Theoretical results for the orbital
magnetic susceptibility in the entire energy range of the 18-band model of
BiTeI at $T = 300$ K.
For (a) we have performed in Eq.~\eqref{eqFukuyama} the $\mathbf
k$~integral over the whole three-dimensional Brioullin zone, whereas for (b) we
have calculated only the contribution from the $ab$ plane, i.\,e., we
have assumed the integrand to be indepent of $k_z$ and taken only the value at
$k_z = \pi/c$.\label{4}}
        \end{center}
\end{figure}

Up to now, we have focused on the experimentally accessible region $-0.15$ eV
$< E_F < 0.15$ eV. It is of interest to study $\chi$ in an extended region to
reveal the mechanism of the enhanced orbital magnetism. Figure~\ref{4}(a) shows
the calculated orbital magnetic susceptibility as a function of~$E_F$ over the
whole energy range of the 18-band model at $T = 300$ K.
Large positive and negative values are
observed, which are even further enhanced if we consider only 
the contribution from the $ab$~plane of the BZ as in Figure~\ref{4}(b).
It turns out that these large dia- and paramagnetic peaks are caused by
(anti)crossings in the band structure of BiTeI,
which can be described as 2D tilted Dirac cones.
A detailed discussion of the appearance of such tilted Dirac cones in
the band structure of BiTeI and their correspondence to the observed dia-
and paramagnetic peaks in Figure~\ref{4} is given in the supplemental material.
There we also provide a simple model which generically describes the dispersion
of tilted band crossings and explains the diverging paramagnetic
orbital susceptibility as the temperature $T$ goes to zero.

In summary, we have studied the orbital magnetism focusing on its enhancement
due to interband effects. As~a concrete example we have considered BiTeI with
its giant Rashba splitting to see how the spin-orbit interaction affects the
orbital magnetism. In addition to the temperature-dependent large diamagnetic
susceptibility near the band crossing point analogous to graphene, we have found
two mechanisms for the enhanced orbital \textit{paramagnetism}. One is the
Rashba splitting with the Fermi energy below the band crossing, and the other is
the tilted band crossing.

This research is supported by MEXT Grand-in-Aid No. 20740167, 19048008,
19048015, and 21244053, Stra\-{}tegic International Cooperative Program (Joint
Research Type) from Japan Science and Technology Agency, and by the Japan
Society for the Promotion of Science (JSPS) through its ``Funding Program for
World-Leading Innovative R \& D on Science and Technology (FIRST Program)''. G.
A. H. S. acknowledges support from MEXT and DAAD.


\begin{thebibliography}{10}
\bibitem{parameters}
In our 2D model, $m^*$ and $\lambda$ are set to $0.09m$ ($m$: electron mass) and
$4.3$ eV\AA, respectively.
\end{thebibliography}
\end{document}